%% file: FA2023_template.tex
\documentclass[10pt]{article}
\usepackage[]{fa2023}
\usepackage{amsmath}
\usepackage{cite}
\usepackage{url}
\usepackage{graphicx}
\usepackage{color}
\usepackage{siunitx}
\usepackage[utf8]{inputenc}

\usepackage{amssymb, subfigure}

\usepackage{tikz}
\usepackage{pgfplots}%
	\usepgfplotslibrary{units}%
	\usepgfplotslibrary{colormaps}%
	\usepgfplotslibrary{groupplots}%
	\usetikzlibrary{positioning}%
	\pgfplotsset{compat=1.9}%
\makeatletter
\pgfplotsset{
 unit code/.code 2 args=
   \begingroup
   \protected@edef\x{\endgroup\si{#2}}\x
}
\usetikzlibrary{patterns}
\pgfplotsset{colormap/myviolet/.style={colormap={myviolet}{rgb255=(0,0,255) color=(white) rgb255=(255,0,0)}}}
\pgfplotsset{colormap/myviolet}
\newcommand{\Hhat}{\widehat{\mathbf{H}}}

\def\NMSE{{\operatorname{NMSE}}}

\title{Implicit neural representation with physics-informed neural networks for the reconstruction of the early part of room impulse responses}
\multauthor
{Mirco Pezzoli$^{*}$ \hspace{1cm} Fabio Antonacci\hspace{1cm} Augusto Sarti} {
  Dipartimento di Elettronica, Informatica e Bioingegneria, Politecnico di Milano, Italy\\
\correspondingauthor{mirco.pezzoli@polimi.it}{Mirco Pezzoli et al.}
}
\begin{document}
\maketitle
\begin{abstract}
Recently deep learning and machine learning approaches have been widely employed for various applications in acoustics. 
Nonetheless, in the area of sound field processing and reconstruction classic methods based on the solutions of wave equation are still widespread.
Recently, physics-informed neural networks have been proposed as a deep learning paradigm for solving partial differential equations which govern physical phenomena, bridging the gap between purely data-driven and model based methods.
Here, we exploit physics-informed neural networks to reconstruct the early part of missing room impulse responses in an uniform linear array. 
This methodology allows us to exploit the underlying law of acoustics, i.e., the wave equation, forcing the neural network to generate physically meaningful solutions given only a limited number of data points.
The results on real measurements show that the proposed model achieves accurate reconstruction and performance in line with respect to state-of-the-art deep-learning and compress sensing techniques while maintaining a lightweight architecture.
\end{abstract}
%
\section{Introduction}\label{sec:introduction}
Sound field reconstruction is fundamental in augmented and virtual reality applications, where users can experience immersive audio environments. 
To accurately characterize the acoustic properties of a given environment, the acquisition of multichannel signals is necessary.
Room impulse responses (RIRs) captured with microphone arrays are particularly useful for this and various tasks such as sound source localization \cite{cobos2017survey,cobos2023acoustic}, separation \cite{gannot2017consolidated,pezzoli2021ray}, and sound field navigation \cite{tylka2020fundamentals,mccormack2022object,pezzoli2020parametric}. 
In fact RIRs provide a model of the sound propagation between the acoustic source and the microphone array within an environment. 

The reconstruction of RIRs or sound field in general, has been a subject of extensive research, leading to the development of two primary categories of solutions: parametric and non-parametric techniques.
Parametric methods \cite{pulkki2018parametric,pezzoli2020parametric,pezzoli2018estimation,pezzoli2018reconstruction,thiergart2013geometry,mccormack2022parametric} rely on simplified parametric models of the sound field to convey an effective spatial audio perception to the user. 
In contrast, non-parametric methods \cite{koyama2019sparse,ribeiro2023kernel,das2021room,antonello2017room,pezzoli2022sparsity} aim to numerically estimate the acoustic field. 
Most of the available techniques in this class are based on compressed sensing principles \cite{donoho2006compressed} combined with the solutions of the wave equation \cite{williams1999fourier}, i.e., plane wave \cite{jin2015theory} and spherical wave \cite{fahim2017sound,pezzoli2022sparsity}, the modal expansion \cite{das2021room} or the equivalent source method (ESM) \cite{antonello2017room,tsunokuni2021spatial}.

A third category comprising deep learning emerged as an alternative approach for sound field reconstruction and a wide range of problems in the field of acoustics \cite{olivieri2021near, bainco2019acousticdeepreview,olivieri2021audio, campagnoli2020vibrational,fernandez2023generative}. 
In \cite{lluis2020sound}, a  convolutional neural network (CNN) has been proposed for the reconstruction of room transfer functions.
However, as noted in \cite{lluis2020sound} the model is limited to low frequencies and the generalization is constrained by the available data set.
To overcome the frequency and data set limitations, in \cite{pezzoli2022deep}, the authors proposed a \textit{deep prior} approach \cite{ulyanov2018deep} to RIR reconstruction in time domain. 
The deep prior paradigm \cite{pezzoli2022deep} considers the structure of a CNN as a regularization prior to learn a mapping from a random input to the reconstructed RIRs of an Uniform Linear Array (ULA).
As a result, no extensive data set is required for training since the optimization is performed over a single ULA.

Recently, in order to exploit the underlying physics of the sound field, a physics-informed neural network (PINN) \cite{raissi2019physics, cuomo2022scientific,olivieri2021physics} for sound field reconstruction has been introduced in \cite{shigemi2022physics}. 
The main idea of PINN \cite{cuomo2022scientific, raissi2019physics} is to force the output of a network to follow the partial differential equations (PDE) governing the system under analysis. 
In particular, PDE computation is performed exploiting the automatic differentiation framework underlying the training procedure of neural networks.
Following the PINN approach, in \cite{shigemi2022physics} the authors augmented the loss function used for training a CNN with the computation of the Helmholtz equation \cite{shigemi2022physics}. 
However, differently from standard PINNs \cite{raissi2019physics}, the network provides as output an estimate of the derivatives required to compute the Helmholtz equation instead of relying on automatic differentiation. 
Moreover, the system works at a fixed frequency (\SI{300}{\hertz}) and it has been tested only on simulated data. 

In this paper, we propose the use of a physics-informed approach for the reconstruction of the early part of RIRs.
As a matter of fact, the early part of RIRs provides relevant information on the geometry of environment \cite{dokmanic2011can, antonacci2012inference} affecting the timbre and localization of acoustic sources \cite{gotoh1977consideration}.
Therefore, accurate reconstruction of the early part of RIRs \cite{tsunokuni2021spatial,alary2020frequency} is required, while the late reverberation is typically modelled through its statistical characteristics \cite{pulkki2018parametric,pezzoli2020parametric,lindau2012perceptual}. 
In order to avoid frequency limitations, we work on RIRs in the time domain.
We adopt a network that takes as input the signal domain i.e., the time and position of the microphone and provides as output an estimate of the RIRs at the given coordinates. 
In order to improve the performance exploiting prior knowledge on the signal domain, we  employed a network structure known as SIREN \cite{sitzmann2020implicit} trained using the PINN paradigm.
We refer to the adopted approach to as physics-informed SIREN (PI-SIREN). 
SIREN demonstrated to be an effective architecture to learn \textit{neural implicit representations} of different signals including audio and for solving the wave equation (direct problem) \cite{sitzmann2020implicit}.
However, the adoption of SIREN has not been fully explored yet for solving time-domain inverse problems in the field of multichannel acoustic processing or applying to real acoustic measurements.
In this work, we investigate the use of PI-SIREN for the reconstruction of early parts of the RIRs acquired by an ULA. 
Results on simulations revealed that in contrast to classical PINN, PI-SIREN is a suitable architecture for RIR reconstruction.
In addition, we compare the reconstructions of PI-SIREN on real data with respect to state-of-the-art solutions based on compressed sensing \cite{zea2019compressed} and deep learning \cite{pezzoli2022deep} showing improved reconstruction of the early parts of the RIRs in two of the three considered rooms.
\section{Problem statement}\label{sec:problem}
\begin{figure}
    \centering
    \input{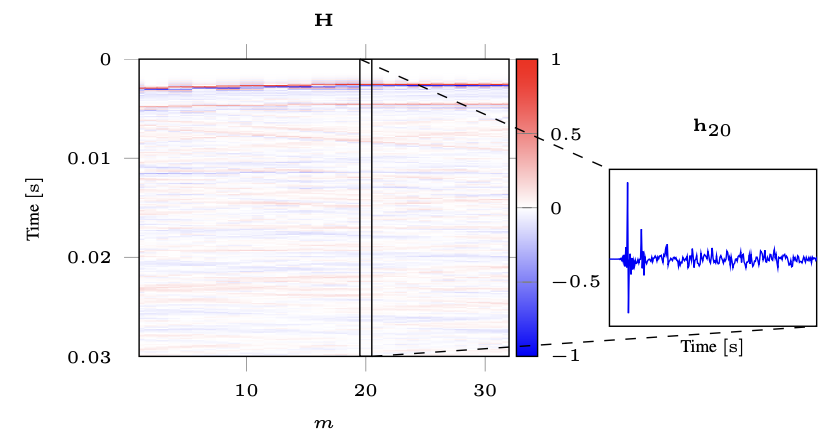}%
    \caption{Example of RIRs $\mathbf{H}$ of a $M=32$ microphones ULA.}%
    \label{fig:rir_image}%
    \vspace{-5em}
\end{figure}
\subsection{RIR data model}
Let us consider an acoustic source located in $\bm{r}'=[x',y',z']^T$ and a set of $M$ microphones acquiring the generated sound field. 
Assuming linear acoustics and absence of noise, the sound pressure at the $m$th sensor can be defined as
\begin{equation}\label{eq:mic_signal}
    p(\bm{r}_m, t) = h(t, \bm{r}_m, \bm{r}') \ast s(t), \quad m=1,\dots,M,
\end{equation}
where $p(\bm{r}_m, t)$ is the time-domain sound pressure at time instant $t$ and location $\bm{r}_m$, $s(t)$ is the signal emitted by the source and $\ast$ denotes the linear convolution operation.
The term $h(t, \bm{r}_m, \bm{r}')$ in \eqref{eq:mic_signal} refers to the RIR between the source in $\bm{r}'$ and the sensor at $\bm{r}_m$. 
In general, RIR provides a description of the sound propagation in the environment from a source to a receiver and due to \eqref{eq:mic_signal}, it completely characterizes the spatial properties of the sound field.
In ideal conditions with unbounded domain, the RIR is given by the well-known Green's function \cite{williams1999fourier} which is a particular solution of the inhomogenous wave equation \cite{williams1999fourier}
\begin{equation}
    \nabla^2p(\bm{r}, t) - \frac{1}{c^2}\frac{\partial^2 p(\bm{r}, t)}{\partial t^2} = \delta(\bm{r}-\bm{r}', t),
\end{equation}
where $\delta$ is the Dirac delta function and $c$ is the speed of sound in air. 

In this work, we consider the RIRs of an ULA and the location of each $m$th microphone is given by the distance $d$ between two consecutive sensors as $r_m = [x_a, (m-1)d, z_a]^T$. 
The values of $x_a$ and $z_a$ are the same for all the sensors in the ULA.
It follows that the maximum frequency for aliasing-free sound field acquisition in the ULA is limited by the distance $d$ through
\begin{equation}\label{eq:freq_aliasing}
    F_{\mathrm{max}} = \frac{c}{2d}.
\end{equation}
In practice, we organize the acquired RIRs in a $N\times M$ matrix defined as
\begin{equation}\label{eq:rir_image}
    \mathbf{H} = [\mathbf{h}_1, \dots, \mathbf{h}_M],
\end{equation}
where $\mathbf{h}_m \in \mathbb{R}^{N\times 1}$ is the vector containing the $N$-length sampled RIR of the $m$th microphone. 
In Fig.~\ref{fig:rir_image}, an example of RIRs acquired by an ULA is shown. 

\subsection{RIR reconstruction problem}
We assume that a limited subset, indexed as $\tilde{\mathcal{M}}$, of the ULA sensors $\mathcal{M}$ is available, and thus $\tilde{\mathcal{M}} \subseteq \mathcal{M}$ $\left(\lvert \tilde{\mathcal{M}} \rvert = \tilde{M} < M\right)$.
The goal of RIR reconstruction is to recover the missing data exploiting the information available from RIRs in the observation points $\left\{\bm{r}_{\tilde{m}}\right\}_{\tilde{m} \in \tilde{\mathcal{M}}}$.
Various techniques have been proposed in the literature to address the spatial-sampling requirement for reconstructing RIRs from an undersampled measurement set. 
In general, this task can be interpreted in the framework of inverse problems, and a solution to the problem can be found through the following minimization
\begin{equation}\label{eq:inverse_problem}
\begin{aligned}
 \bm{\theta}^* &= \underset{{\bm{\theta}}}{\text{argmin}}\,\,J\left(\bm{\theta}\right) = \\ &E\left(f(\bm{\theta}, \left\{\bm{r}_{\tilde{m}}\right\}_{\tilde{m} \in \tilde{\mathcal{M}}} ) - \mathbf{H}(\left\{\bm{r}_{\tilde{m}}\right\}_{\tilde{m} \in \tilde{\mathcal{M}}}))\right),
\end{aligned}
\end{equation}
where $E(\cdot)$ is a data-fidelity term, e.g., the mean squared error, between the estimated data and the observations, and $f(\bm{\theta}, \bm{r})$ is a function that generates the estimated RIRs using the parameters $\bm{\theta}$.
The time dependency in \eqref{eq:inverse_problem} has been omitted for notational simplicity.
It is worth noting that in \eqref{eq:inverse_problem}, the evaluation of $f$ is performed in the observation locations $\left\{\bm{r}_{\tilde{m}}\right\}_{\tilde{m} \in \tilde{\mathcal{M}}}$. 
However, $f$ must be able to provide a meaningful estimate also in location that are different from the available ones. 
Therefore, the solution to the ill-posed problem \eqref{eq:inverse_problem} is constrained using regularization strategies. 
Typical techniques include compressed sensing frameworks based on assumptions about the signal model such as plane and spherical wave expansions \cite{koyama2019sparse}, ESM \cite{tsunokuni2021spatial}, or the RIRs structure \cite{zea2019compressed}, as well as deep learning approaches \cite{lluis2020sound,pezzoli2022deep}.
\section{Proposed Method}
In this work, we aim at solving the RIR reconstruction problem \eqref{eq:inverse_problem} in order to provide an estimate of the ULA RIRs as
\begin{equation}
    \Hhat = f\left(\bm{\theta}^*, \left\{\bm{r}_{{m}}\right\}_{{m} \in {\mathcal{M}}}\right),
\end{equation}
where the function $f(\cdot)$ represents a neural network. 
In particular, we adopt the structure of a SIREN \cite{sitzmann2020implicit} neural network. 
SIREN proved to be an effective architecture for learning the so-called \textit{neural implicit representations} of different classes of signals, including audio signals. 
The proposed model has the structure of a multilayer perceptron (MLP) with sinusoidal activation functions, for which the $i$th layer can be expressed as
\begin{equation}\label{eq:sine_layer}
    \phi_i(\mathbf{x}_i) = \sin\left(\omega_0 \mathbf{x}_i^T \bm{\theta}_i + \mathbf{b}_i\right), 
\end{equation}
where $\mathbf{x}_i$, $\bm{\theta}_i$, and $\mathbf{b}_i$ are the input vector, the weights and the biases of the $i$th layer, respectively, while $\omega_0$ is an initialization hyper-parameter \cite{sitzmann2020implicit}.
The adopted SIREN architecture is thus a composition of $L$ layers 
\begin{equation}
    f\left(\bm{\theta}, \mathbf{x}\right) = \left(\phi_L \circ \phi_{L-1}, \dots, \phi_1 \right)(\mathbf{x}),
\end{equation}
where $\mathbf{x}$ is the input of the network while $\bm{\theta}$ is the set of learnable parameters. 
Following the paradigm of neural implicit representations, the SIREN model takes as input the signal domain, namely the sensor position $\bm{r}_m$ and the time instant $t$ and provides as output an estimate of the RIR $\hat{h}(t, \bm{r}_m)$. 
Hence, the role of the network is to provide a parameterized representation of the signals through the parameters of the MLP.
Essentially, during the training, the neural network overfits the available signals becoming an alternative implicit representation of the RIRs.

Although we can fit the available RIRs through SIREN, regularization strategies are required in order to provide meaningful results in different points of the domain i.e., to estimate the missing RIRs.

Here, we consider training SIREN using the PINN approach, denoting the solution as PI-SIREN. 
Using as the target for the training the reconstruction of the observation only, there is no guarantee that the solution follows the physical law of the underlining problem, namely the wave equation \cite{williams1999fourier}. 
PINN are forced to learn solutions that follows the PDE of the underlying physics in order to obtain improved results. 
This approach exploits the prior knowledge on the system in order to regularize the estimation of the neural network. 
Therefore, we adopted the following loss function for training PI-SIREN which includes a physics-informed term as
{\begin{equation}\label{eq:loss_function}
\begin{aligned}
    \mathcal{L} &= \frac{1}{\Tilde{M}} \sum_{\tilde{m}\in \tilde{\mathcal{M}}} \lVert \hat{{h}}_{\tilde{m}} -  {h}_{\tilde{m}}\rVert^2_2 + \\
    &\lambda \frac{1}{{M}} \sum_{m=1}^{M} \left\lVert \frac{1}{c^2} \frac{\partial^2\hat{{h}}_m}{\partial t^2} - \nabla^2 \hat{h}_m \right\rVert_2^2,
\end{aligned}
\end{equation}}
where $\lVert \cdot \rVert_2$ is the $\ell_2$ norm, the first term of the summation represents a distance between the prediction and the available data, while the second term corresponds to the PDE loss given by the wave equation and weighted by parameter $\lambda$.
While the first part of \eqref{eq:loss_function} makes the network fit the observation, the PDE term constraints the output to follow the wave equation.
The use of the PDE loss results in a regularized solution since the output conforms with the underlying physical equation.
Once trained, PI-SIREN can be used to obtain the RIRs at the missing and available positions of the ULA simply feeding the network with the locations $\bm{r}_m$, $m=1,\dots,M$, and the different time instants $t$.

\begin{figure*}
    \centering
    \begin{subfigure}[]{%
    \includegraphics[width=0.4\columnwidth]{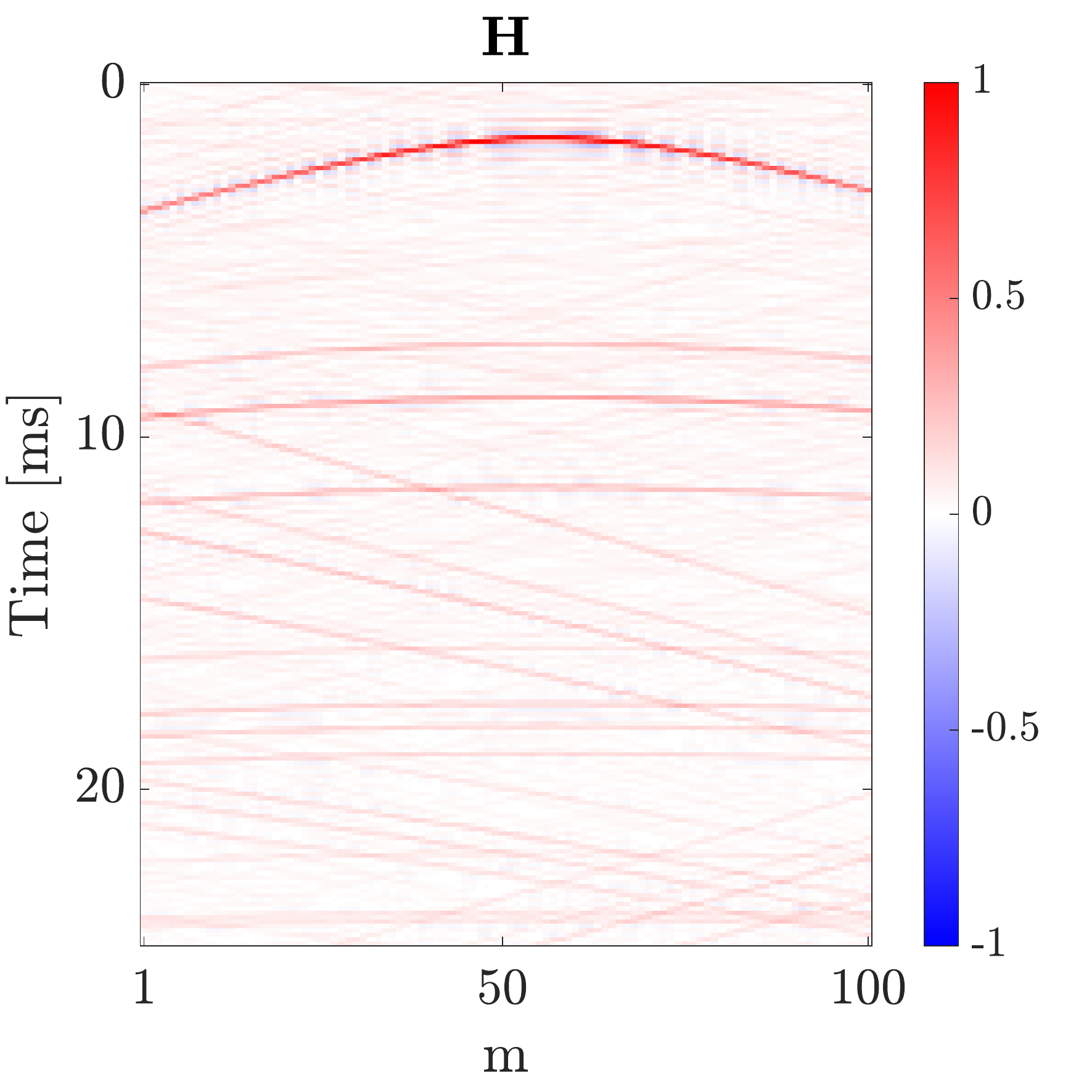}\label{fig:gt}}%
    \end{subfigure}%
    \begin{subfigure}[]{%
    \includegraphics[width=0.4\columnwidth]{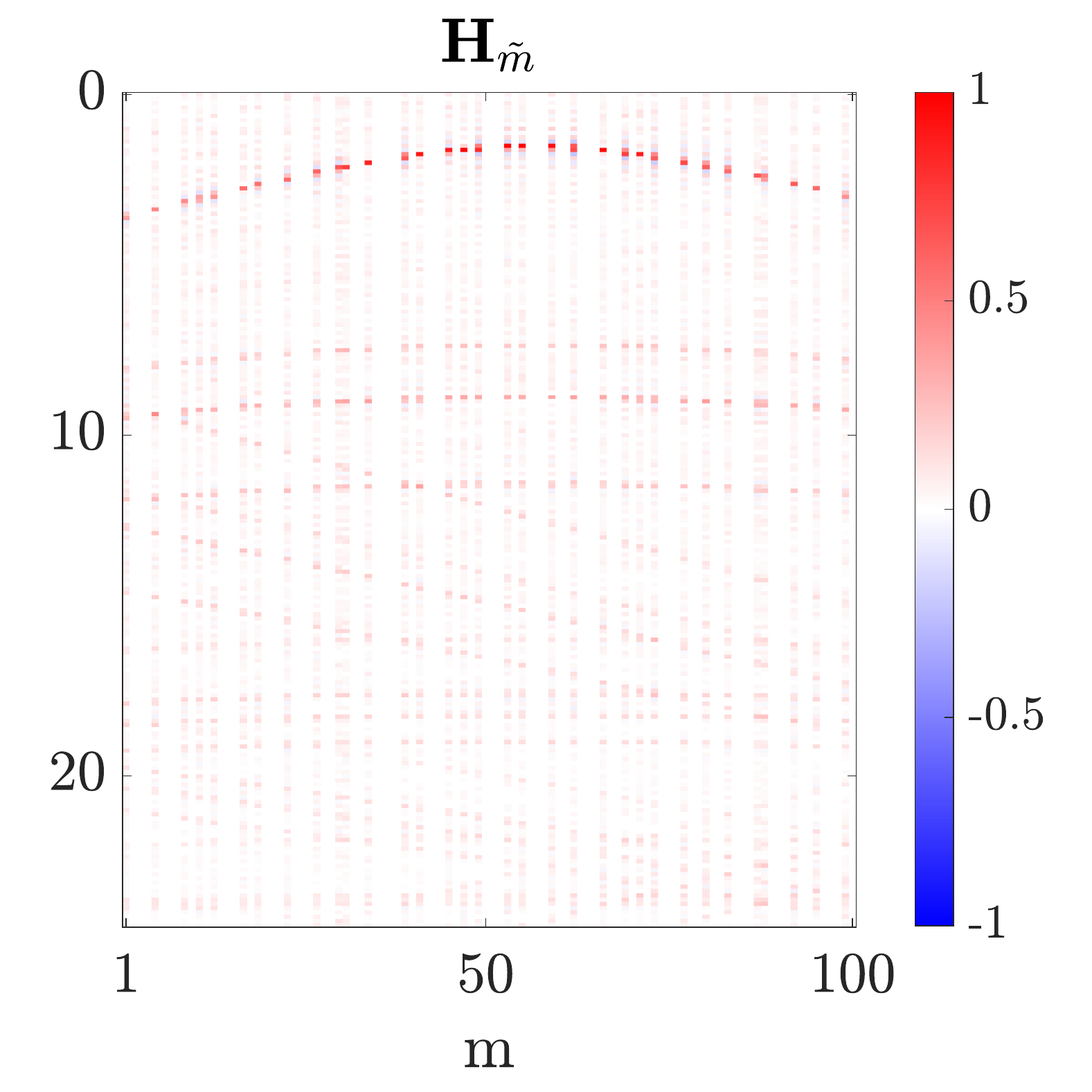}\label{fig:bs}}%
    \end{subfigure}
    \begin{subfigure}[]{%
    \includegraphics[width=0.4\columnwidth]{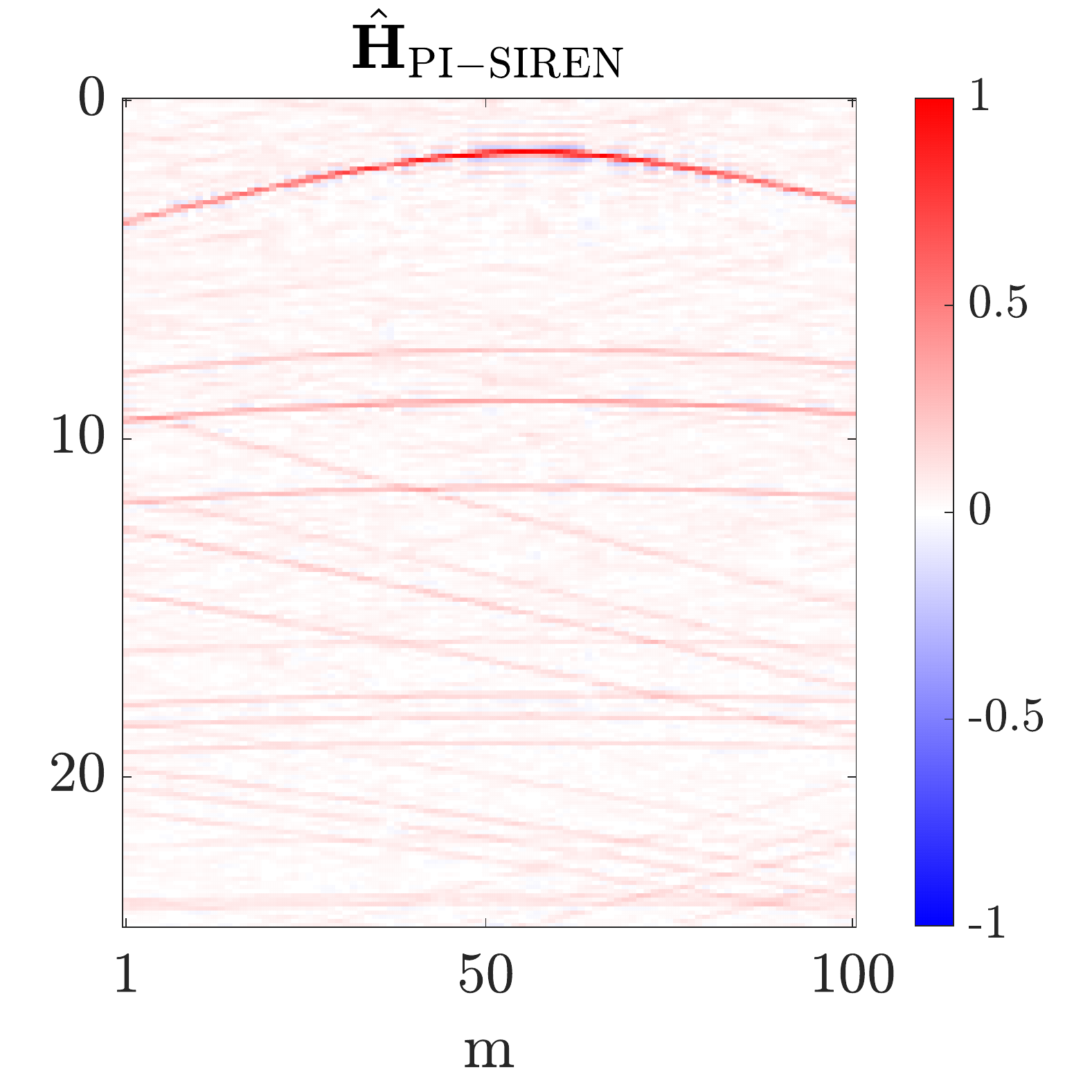}\label{fig:pi-siren}}%
    \end{subfigure}
    \begin{subfigure}[]{%
    \includegraphics[width=0.4\columnwidth]{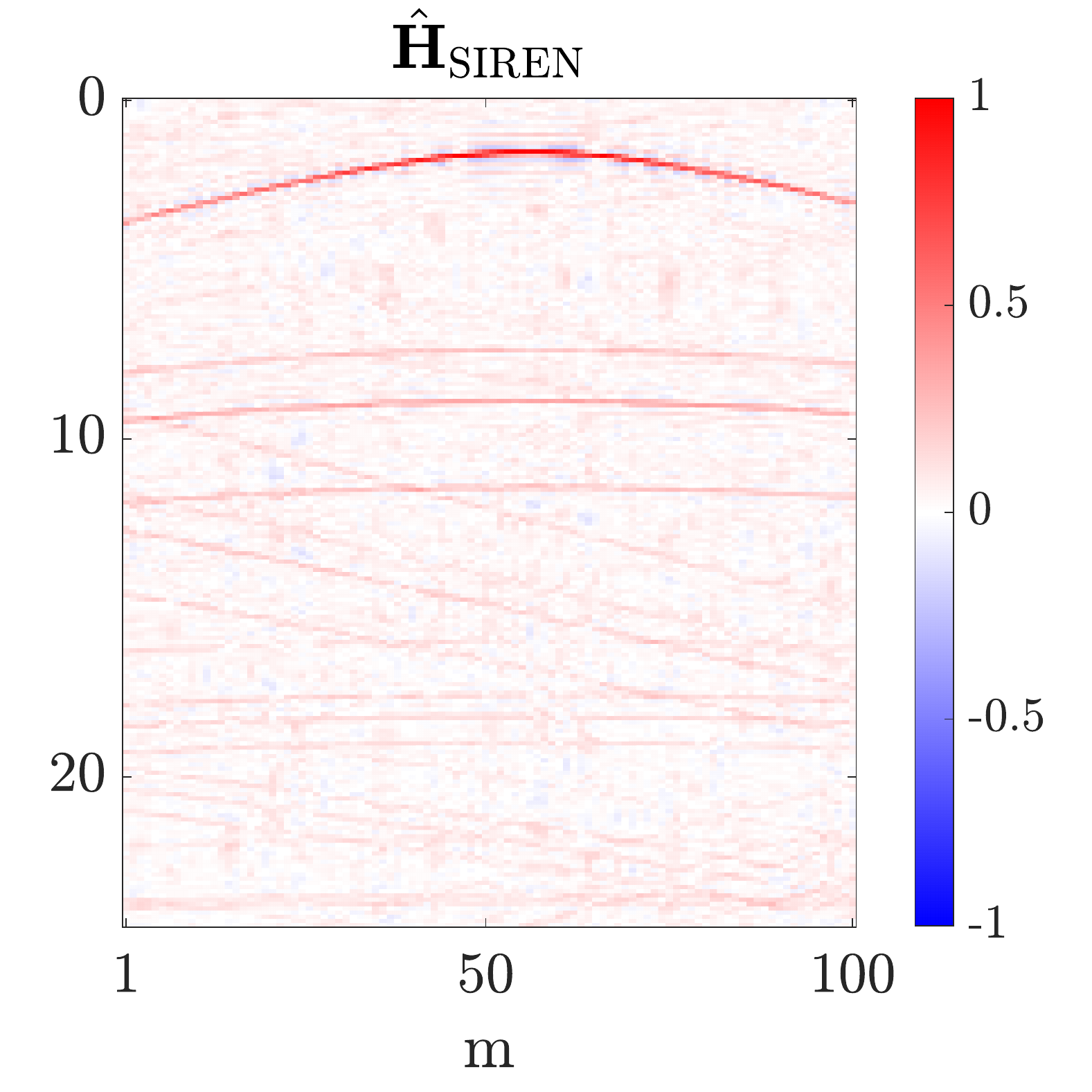}\label{fig:siren}}%
    \end{subfigure}
     \begin{subfigure}[]{%
    \includegraphics[width=0.4\columnwidth]{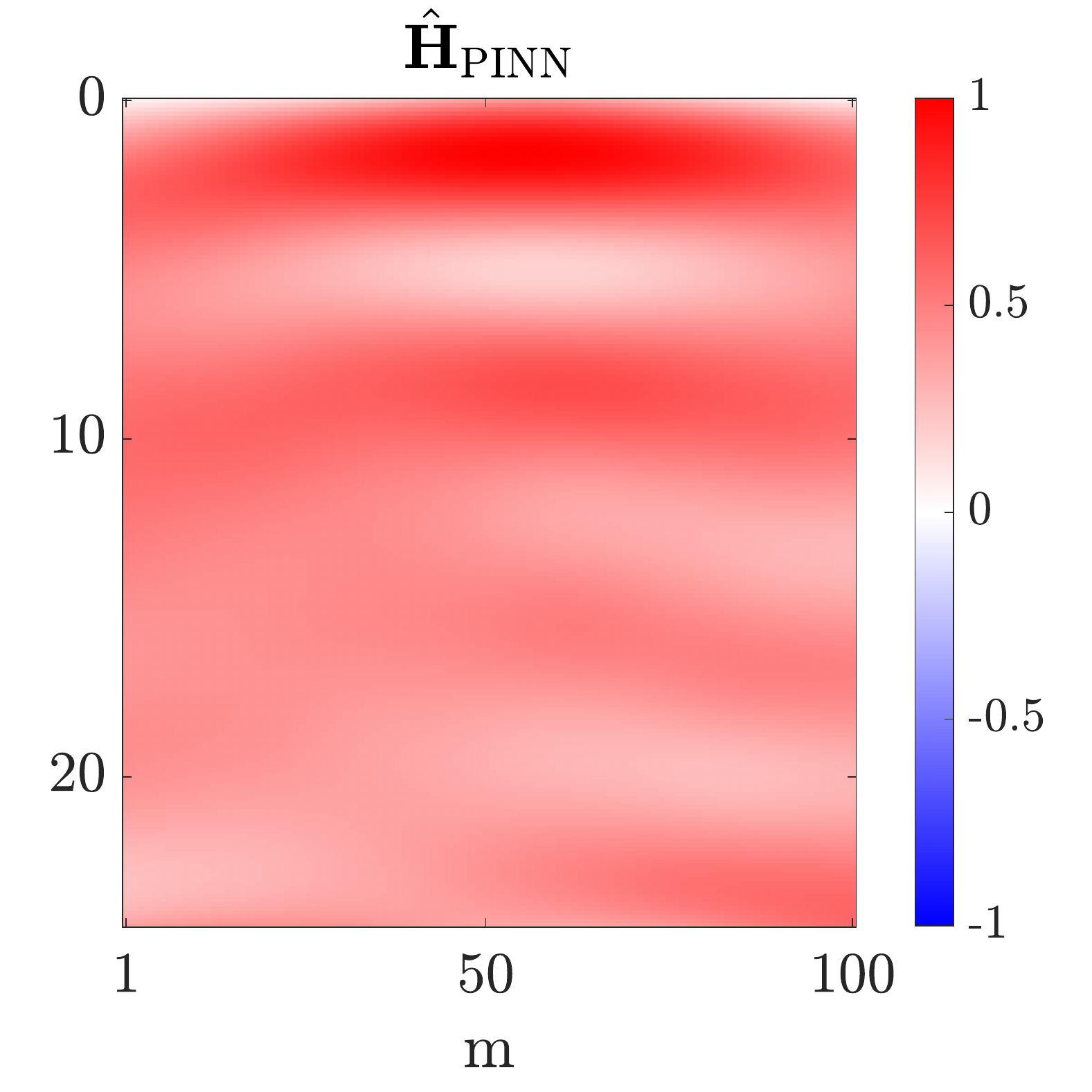}\label{fig:pinn}}%
    \end{subfigure}
    \caption{(a) Simulated RIRs $\mathbf{H}$. (b) The observation $\mathbf{H}_{\tilde{m}}$ of $\tilde{M}=33$ microphones employed as input for the networks. The reconstructions obtained using PI-SIREN (c), SIREN (d) and PINN (e).}
    \label{fig:simulation}
\end{figure*}
\section{Numerical experiments}
\subsection{Setup}
We evaluate the performance of PI-SIREN for RIR reconstruction on both simulated and measured data from \cite{zea2019compressed}. 
We considered an ULA of $M=100$ microphones with distance $d=\SI{2.02}{\centi\meter}$ which gives a maximum frequency \eqref{eq:freq_aliasing} $F_{\mathrm{max}} = \SI{8.489}{\kilo\hertz}$. 
The simulated RIRs have been computed at sampling rate $\SI{8}{\kilo\hertz}$ using the image source technique \cite{habets2006rirgenerator} for a shoe-box room of dimensions $\SI{6}{\meter}$ $\times$ $\SI{4}{\meter}$ $\times$ $\SI{3}{\meter}$ and reverberation time $\mathrm{T}_{60}$ $=$ $\SI{0.5}{\second}$.
For this work we limit the analysis to the first $\SI{20}{\milli\second}$ of the RIRs which corresponds to the early part (direct and early reflections) of the impulse response.

The proposed PI-SIREN architecture is composed of $L=5$ layers of $256$ neurons in which the last layer is linear. 
The network has a total of $198401$ trainable parameters.
The initialization frequency $\omega_0$ in \eqref{eq:sine_layer} is set to $15$ for the first layer, while as in \cite{sitzmann2020implicit} $\omega_0=30$ for the hidden layers.
The network is trained for 2000 iterations using Adam optimizer with learning rate equal to $10^{-4}$. 
The weight parameter in \eqref{eq:loss_function} has been experimentally set to $\lambda=5\cdot10^{-15}$.
Similarly to \cite{pezzoli2022deep,zea2019compressed}, we evaluate the reconstruction performance in terms of the normalized mean square error ($\NMSE$) between the reconstructed data and the reference RIRs defined as \cite{zea2019compressed}
\begin{equation}\label{eq:nmse}
    \NMSE\left(\Hhat, \mathbf{H}\right) = 10\log_{10} \frac{1}{M} \sum_{m=1}^{M} \frac{\lVert \widehat{\mathbf{h}}_m - \mathbf{h}_m \rVert^2}{\lVert \mathbf{h}_m \rVert^2},
\end{equation}
where $\widehat{\mathbf{h}}_m$ is the $m$th RIR estimate provided by the reconstruction technique. 
The observations are computed considering $\Tilde{M} = \left\{20,33\right\}$ microphones randomly selected as in \cite{zea2019compressed}, which corresponds to  ${1}/{5}$ and $1/3$ of sensors available, respectively. 
\subsection{Results}
In order to assess the effectiveness of the proposed PI-SIREN methodology in terms of architecure and training strategy, we evaluate the reconstruction performance on simulated data. 
We compare the reconstruction of PI-SIREN with respect to a classical PINN architecture \cite{raissi2019physics} and SIREN trained without the wave equation term in the loss function \eqref{eq:loss_function}. 
The PINN shares the same structure of PI-SIREN in terms of layers and parameters, however $\tanh$ is adopted as nonlinear function of the neurons. 
In Fig.~\ref{fig:simulation}, the RIRs $\mathbf{H}$ along with the observation $\mathbf{H}_{\tilde{m}}$ with $\tilde{M} = 33$ and the obtained reconstructions are reported.

From Fig.~\ref{fig:pinn}, we can observe that PINN fails to reconstruct the RIRs, obtaining a $\NMSE_{\mathrm{PINN}}$ $=$ $\SI{14.5}{\decibel}$. 
The adoption of the sinusoidal activation function in SIREN (see Fig.~\ref{fig:siren}) determines an improved reconstruction performance with $\NMSE_{\mathrm{SIREN}}$ $=$ $\SI{-7.1}{\decibel}$. 
Inspecting Fig.~\ref{fig:siren}, we can observe that SIREN reconstructs the direct path and the early reflections in $\hat{\mathbf{H}}_{\mathrm{SIREN}}$, filling the missing channels. 
It follows that SIREN provides an effective implicit representation of the considered signals thanks to the use of the sinusoidal nonlinearity. 
In \cite{pistilli2022signal}, the authors show how a two-layers SIREN can be related to a discrete cosine transform (DCT) of the signal. 
In the context of this work, the consideration in \cite{pistilli2022signal} can be loosely interpreted in terms of a real-valued plenacoustic representation \cite{ajdler2006plenacoustic} of the RIRs. 
Nonetheless, the reconstruction in Fig.~\ref{fig:siren} contains noisy components and the estimated wave fronts at some of the missing locations are incoherent. 
In Fig.~\ref{fig:pi-siren}, the output of PI-SIREN is depicted. 
It is possible to note that, differently from the basic SIREN, PI-SIREN is able to estimate the RIRs more accurately, coherently reconstructing the wave fronts at the missing locations.
The reconstruction of PI-SIREN achieves a $\NMSE_{\mathrm{PI-SIREN}}=\SI{-11.2}{\decibel}$ which is lower with respect to both SIREN and the PINN.
Through the physics-informed loss function in \eqref{eq:loss_function}, in fact, the output of the network is forced to conform with the physical prior of the wave equation.
Therefore, the adoption of the physics-informed loss function in PI-SIREN allows us to obtain an improved performance. 
\begin{table}[t]
    \label{tab:zea_results}
    \resizebox{\columnwidth}{!}{%
    \centering{
    \newcolumntype{?}{!{\vrule width 1pt}}
    \begin{tabular}{c?c|c?c|c?c|c?c|c?c|c?c|c?c|c?c|c?c|c?}
         Room & \multicolumn{2}{c?}{Balder} & \multicolumn{2}{c?}{Freja} & \multicolumn{2}{c?}{Munin} \\
         Mic. & $20$ & $33$ & $20$ & $33$ & $20$ & $33$\\
        \hline
        \multicolumn{7}{c?}{$\NMSE$ $[\si{\decibel}]$}\\
        \hline
        CS  & 
        -5.87 & -11.47 & 
        \textbf{-5.89 }& \textbf{-11.01} & 
        -7.52 & -15.25\\
        DP  & 
        -5.52 & -11.44 & 
        -4.68 & -9.21 & 
        -8.98 & -16.03\\
        PI-SIREN  & 
        \textbf{-6.26} & \textbf{-11.74} & 
        -5.65 & -10.61 & 
        \textbf{-10.00} & \textbf{-16.17}\\
        \hline
    \end{tabular}}}
    \caption{$\NMSE$ of the considered techniques at different downsampling conditions for the three rooms.}
\end{table}
\begin{figure*}
    \centering
    \begin{subfigure}[]{%
    \includegraphics[width=0.5\columnwidth]{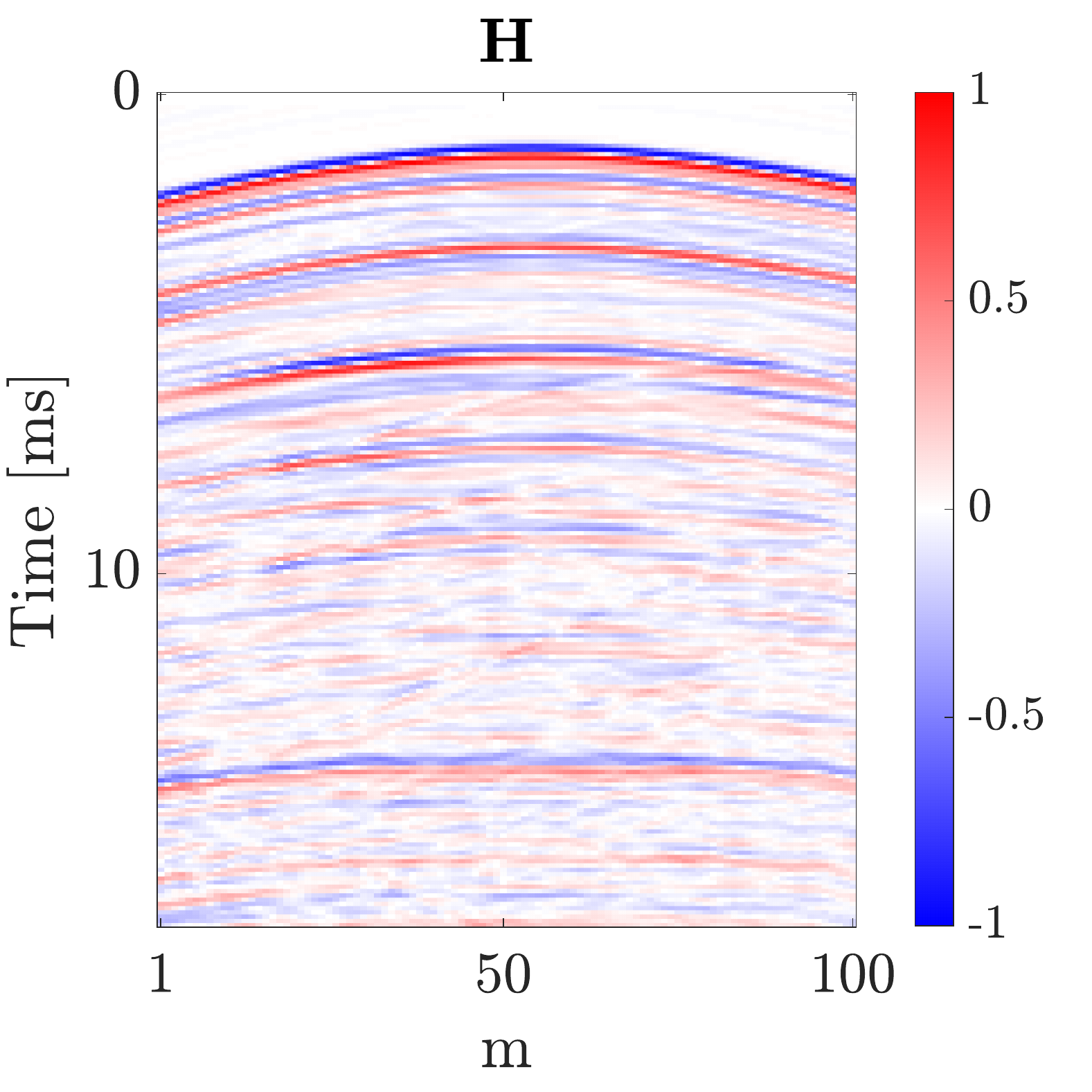}\label{fig:munin_gt}}%
    \end{subfigure}%
    \begin{subfigure}[]{%
    \includegraphics[width=0.5\columnwidth]{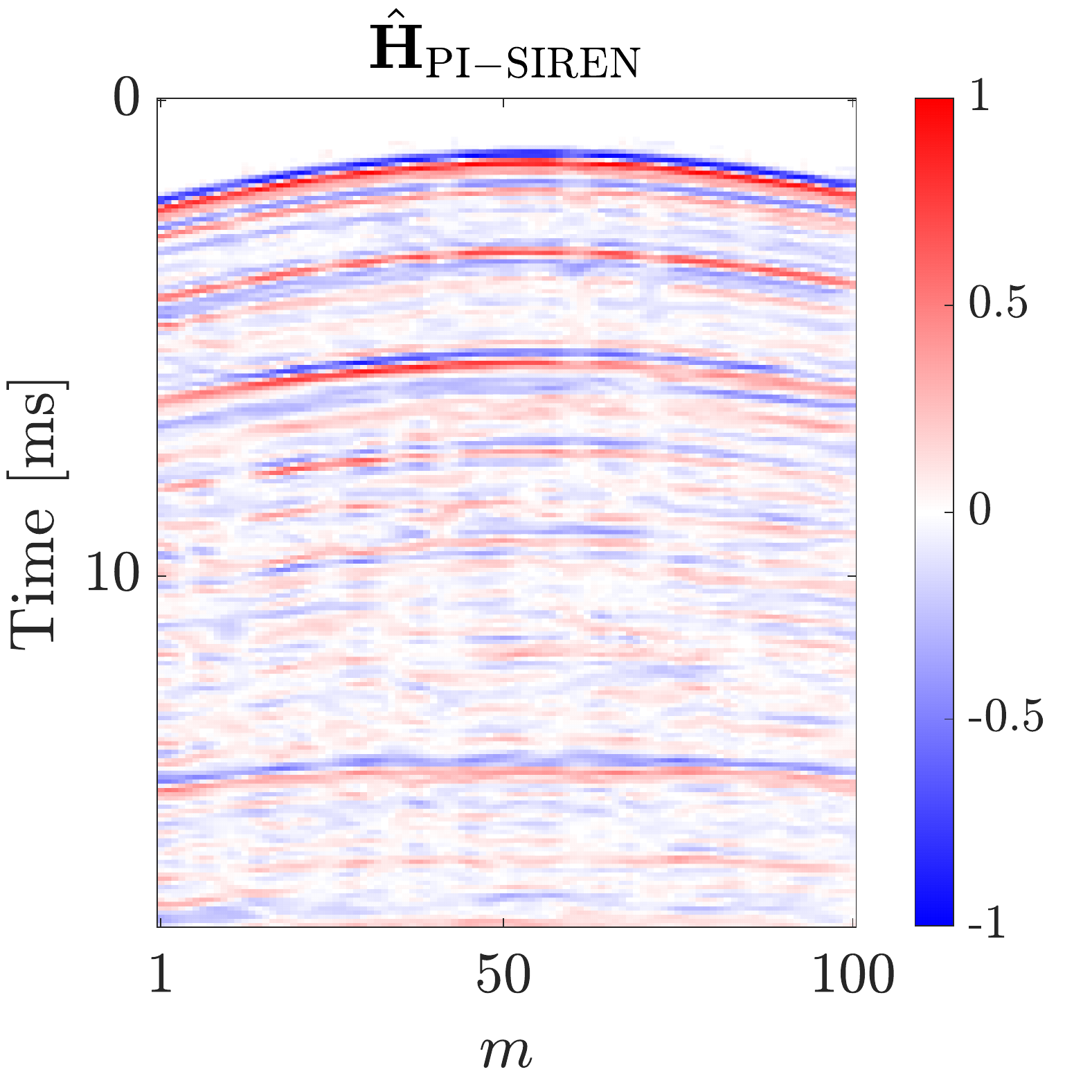}\label{fig:munin-siren}}%
    \end{subfigure}
    \begin{subfigure}[]{%
    \includegraphics[width=0.5\columnwidth]{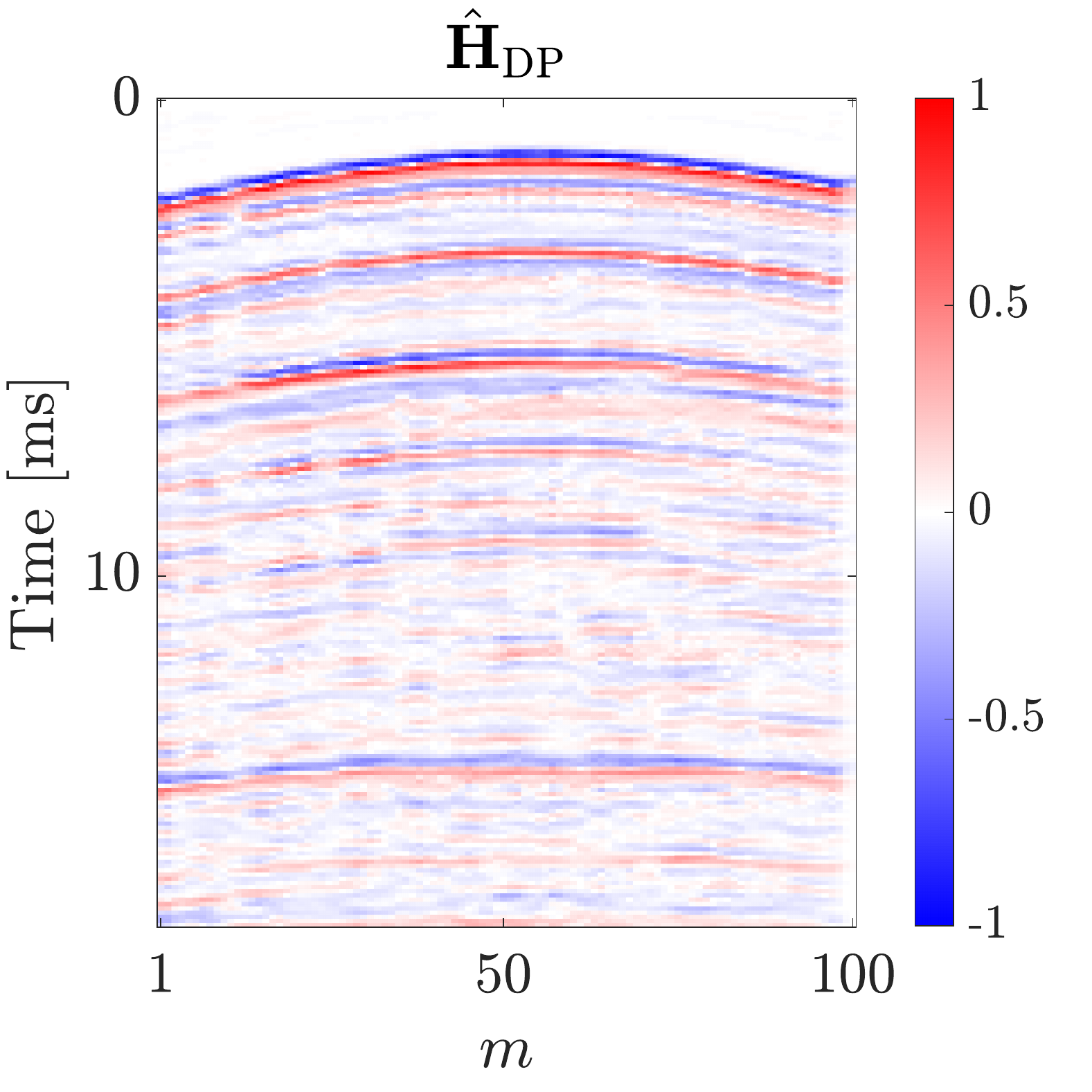}\label{fig:munin-dp}}%
    \end{subfigure}
    \begin{subfigure}[]{%
    \includegraphics[width=0.5\columnwidth]{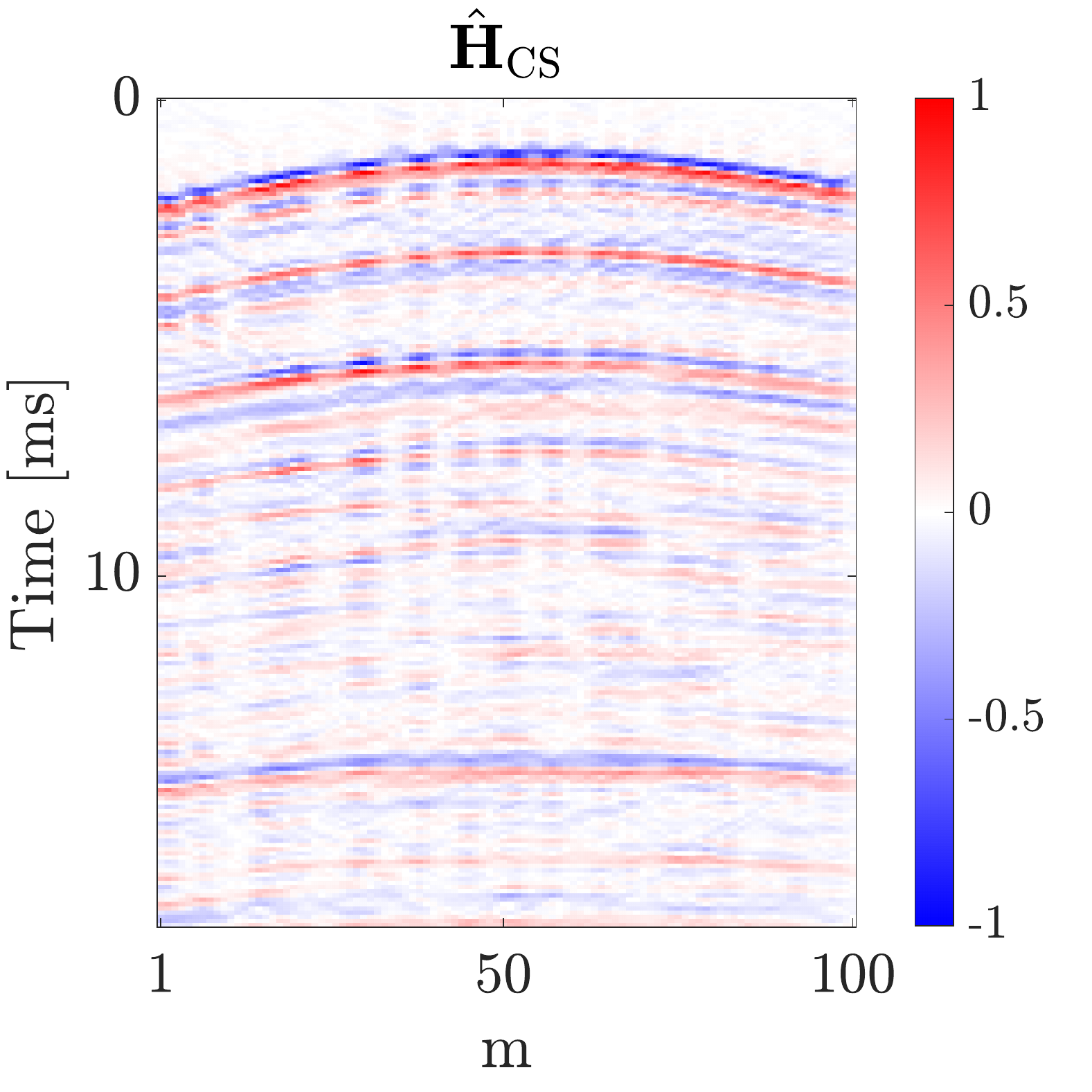}\label{fig:munin-zea}}%
    \end{subfigure}
    \caption{Reconstruction of the RIRs of Munin room using $\tilde{M}=20$ available sensors. (a) The measured RIRs $\mathbf{H}$. The reconstructions are obtained using the proposed model $\hat{\mathbf{H}}_{\mathrm{PI-SIREN}}$ (b), DP \cite{pezzoli2022deep} (c) and CS \cite{zea2019compressed} (d).}
    \label{fig:zea}
\end{figure*}
\subsection{Experimental results}
We evaluate the performance of PI-SIREN on real RIRs measured in three rooms \cite{zea2019compressed} and we compare the estimated reconstruction with respect to the compressed sensing method (CS) in \cite{zea2019compressed} and the deep prior (DP) methodology of \cite{pezzoli2022deep}.
The employed ULA consists of $M=100$ sensors with distance $d=\SI{3}{\centi\meter}$. 
The rooms are named ``Balder'', ``Freja'' and ``Munin'' and the estimated reverberation times $\mathrm{T}_{30}$ are $\SI{0.32}{\second}$, $\SI{0.46}{\second}$ and $\SI{0.63}{\second}$, respectively.

In Table~\ref{tab:zea_results}, the $\NMSE$ obtained for the different rooms are reported. 
As expected, when a lower number of sensors $\tilde{M}=20$ is available the reconstruction performance is reduced for all the considered techniques. 
The performance of the three methods is in line for all the considered scenarios. 
However, the proposed PI-SIREN is able to achieve lower $\NMSE$ in Balder and Munin rooms for both the adopted undersampling conditions.
Interestingly, the best reconstruction performance is achieved for room Munin for every method. 
This room has the highest $\mathrm{T}_{30}$, but as notices in \cite{zea2019compressed}, the density of early reflections is lower with respect to the other rooms, making the reconstruction less challenging. 
CS obtained the lowest $\NMSE$ in room Freja. 
However, the difference with respect of the proposed model is limited to $\SI{0.24}{\decibel}$ and $\SI{0.4}{\decibel}$ for the $\tilde{M}=20$ and $\tilde{M}=33$ scenarios, respectively. 

In Fig.~\ref{fig:zea}, the reconstructions of $\mathbf{H}$ for room Munin given $\tilde{M}=20$ microphones are reported. 
The reference RIRs are depicted in Fig.~\ref{fig:munin_gt}. 
Inspecting the reconstruction in Fig.~\ref{fig:zea}, we can note that all the three methods managed to reconstruct the main structure of the RIRs. 
However, the reconstruction provided by CS presents an underestimation of the RIRs at the missing locations which are seen as light vertical stripes in Fig.~\ref{fig:munin-zea}.
Instead $\mathbf{H}_{\mathrm{PI-SIREN}}$ and $\mathbf{H}_{\mathrm{DP}}$ have a similar performance with a lower reconstruction error ($\NMSE_{\mathrm{PI-SIREN}}=\SI{-10}{\decibel}$) for the proposed model compared to DP ($\NMSE_{\mathrm{DP}}=\SI{-8.89}{\decibel}$). 
\section{Conclusion}
In this work we proposed the use of PINNs for the reconstruction of early part of RIRs. 
The devised architecture consists of a SIREN neural network trained exploiting the physics-informed neural network framework.
This allows us to impose the governing wave equation to the solution of the RIR reconstruction. 
The results shows that the SIREN architecture itself provides an implicit representation of the data.
Moreover, the adoption of the physics-informed training demonstrated  to improve the reconstruction performance.
We investigated the application of the proposed model on real data, showing competitive results with state-of-the-art techniques based on compressed sensing and deep learning. 
The proposed technique is appealing since it synergestically exploits the flexibility of deep learning and the prior knowledge of physics. 
We foresee the future of this work concerning the network design and the modeling of the whole RIRs that can improve the performance and the applicability with respect to the current results.
\section{Acknowledgments}
This work has been funded by "REPERTORIUM project. Grant agreement number 101095065. Horizon Europe. Cluster II. Culture, Creativity and Inclusive society. Call HORIZON-CL2-2022-HERITAGE-01-02."
\bibliography{fa2023_template}
%
%
%
%

\end{document}

%% file: rir_image.tex
\pgfplotsset{%
yticklabel style={
        /pgf/number format/fixed,
        /pgf/number format/precision=5
},%
scaled y ticks=false}%
\begin{tikzpicture}%
\begin{axis}[%
            name=rirImg,
            width = 0.65\columnwidth,
            xlabel=$m$,
            ylabel=Time,
            y unit=\second,
            label style={font=\tiny},
            colorbar,
            point meta min=-1,
            point meta max=1,
            colorbar style={
                at={(1.02,1.0)},
                label style={font=\tiny},
                colormap/myviolet,
                width=6pt
            },
            every tick label/.append style={font=\tiny},
            point meta= explicit, 
            tick align=outside,
            title={\tiny$\mathbf{H}$},
            axis on top,
            colormap/myviolet,
            xmax=32,
            xmin=1,
            ymax=0.03,
            ymin=0,
            y dir = reverse
        ]
        \addplot [matrix plot*,point meta=explicit,mesh/cols=32] file {./images/data/rir_image.txt};
        \coordinate (c1) at (axis cs:19.5,0);
        \coordinate (c2) at (axis cs:20.5,0.03);
        \draw (c1) rectangle (c2);
\end{axis}
\begin{axis}[
        name=zoomRIR,
        at={($(rirImg.south east)+(1cm,0.3cm)$)},
        title={\tiny$\mathbf{h}_{20}$},
        width=0.45\columnwidth,
        every tick label/.append style={font=\tiny},
        point meta min=-1,
        point meta max=1,
        point meta= explicit, 
        xmin = 0, 
        xmax= 0.03,
        xlabel=Time, 
        label style={font=\tiny},
        x unit = \second,
        xtick = \empty,
        ytick = \empty,
         ]
        \addplot[blue] file {images/data/rir.txt};
\end{axis}  
    \draw [dashed] (c2) -- (zoomRIR.south east);
    \draw [dashed] (c1) -- (zoomRIR.north west);
\end{tikzpicture}